\documentclass[final]{IEEEtran}
\ifCLASSINFOpdf
\else
\fi

\usepackage{amsthm,amssymb,graphicx,multirow,amsmath,bm,,color,amsfonts}%,ulem}
\usepackage[update,prepend]{epstopdf}
\usepackage{cite}
\usepackage{tabulary}
\usepackage{bbm} % for \mathbbm{1}
\usepackage{multirow}
\usepackage{comment}
\usepackage[caption=false]{subfig}
\usepackage[ruled,linesnumbered]{algorithm2e}
\SetKw{KwBy}{by}
\usepackage[normalem]{ulem}

\makeatletter
\newcommand{\nosemic}{\renewcommand{\@endalgocfline}{\relax}}% Drop semi-colon ;
\newcommand{\dosemic}{\renewcommand{\@endalgocfline}{\algocf@endline}}% Reinstate semi-colon ;
\let\oldnl\nl% Store \nl in \oldnl
\newcommand{\nonl}{\renewcommand{\nl}{\let\nl\oldnl}}% Remove line number for one line
\usepackage{makecell}

\usepackage{optidef}

\def\nb0{{\mathbf{0}}}
\def\nb1{{\mathbf{1}}}

\newtheorem{lemma}{Lemma}

\def\figref#1{Fig.\,\ref{#1}}%

\DeclareMathOperator{\Tr}{Tr}
\DeclareMathOperator{\rank}{rank}

\allowdisplaybreaks % Allows breaking of eqnarray over multiple pages (avoids unnecessary blanks in the document before eqnarray)
\begin{document}
\pagenumbering{gobble}
\graphicspath{{./figures/}}
\title{
	High Altitude Platform-Based Caching and Multicasting for Rural Connectivity
}
\author{
	Yongqiang Zhang, \IEEEmembership{Graduate Student Member,~IEEE,} Mustafa A. Kishk, \IEEEmembership{Member,~IEEE,} and Mohamed-Slim~Alouini, \IEEEmembership{Fellow,~IEEE}
	\thanks{Yongqiang Zhang and Mohamed-Slim Alouini are with Computer, Electrical and Mathematical Science
		and Engineering Division, King Abdullah University of Science and Technology (KAUST), Thuwal 23955-6900, Kingdom of Saudi Arabia. Email: \{yongqiang.zhang.2, slim.alouini\}@kaust.edu.sa.}
	\thanks{Mustafa A. Kishk is with the Department of Electronic Engineering, Maynooth University, Maynooth, W23 F2H6, Ireland. Email: mustafa.kishk@mu.ie.}
}

\maketitle

\begin{abstract}
	Providing efficient and reliable content delivery in rural areas remains a significant challenge due to the lack of communication infrastructure. To bridge the digital divide, this paper investigates the potential of leveraging multiple high-altitude platforms (HAPs) for energy-efficient content delivery in wide rural regions. Each caching-enabled HAP is equipped with both Free-Space Optical (FSO) transceivers for backhaul links and Radio Frequency (RF) antenna arrays for access links. To further enhance network efficiency, we consider a network coding-based multicasting scheme, where different types of content are treated as distinct multicast sessions.
	 With the objective of minimizing long-term power cost, we propose a hierarchical framework that integrates deep reinforcement learning (DRL)
	  and convex optimization to jointly optimize dynamic caching strategies and resource allocation across the network. Simulation results demonstrate that our approach significantly reduces power cost compared to several baseline approaches, providing a practical solution for improving rural connectivity.
\end{abstract}

\section{Introduction}\label{sec:intro}
% Connectivity in Rural
Due to the absence of adequate communication resources, providing consistent and high-quality network access to rural and remote areas presents a significant challenge \cite{Imran2024}.
While successful in urban areas, many communication technologies cannot be directly applied to underserved regions due to limited resources such as FSO, cellular network infrastructure, and reliable power supply \cite{Chiaraviglio2017}.
In 6G and future networks, non-terrestrial networks (NTNs), including HAPs,  low altitude platforms (LAPs), and satellites, have emerged as promising alternatives to extend connectivity to these regions \cite{Giordani2021}. 
Compared to LAPs, such as unmanned aerial vehicles (UAVs), HAPs exhibit longer operational endurance and can integrate advanced communication systems due to their greater payload capacity \cite{KarabulutKurt2021}.
While satellites provide broader coverage, their continuous movement restricts sustained service to specific areas, making HAPs more effective in maintaining long-term connectivity for ground users \cite{Arum2020}.

Integrated access and backhaul (IAB) has been recognized by 3rd generation partnership project (3GPP) as a cost-effective solution for backhaul by allocating a portion of radio resources for wireless backhauling, thereby eliminating the need for expensive fiber-optic cable deployment \cite{Polese2020}.
Without spectrum licensing requirements, FSO is an attractive cost-effective communication technology that uses optical signals to provide high-speed backhaul transmission, particularly in scenarios where the deployment of traditional fiber-optic-based backhaul is either too costly or impractical \cite{Alzenad2018}. 
Recent works demonstrate that innovative hybrid modulation and multiplexing techniques greatly enhance the capacity and robustness of FSO backhaul links under atmospheric impairments \cite{Elsayed2024a,Elsayed2024}.
Owing to its larger carrying capability, a HAP can affordably accommodate both a multi-antenna RF transmitter and an FSO transceiver simultaneously \cite{Alam2021}.
Leveraging this functionality, a HAP can support FSO backhaul connections to other HAPs and offer RF access links to terrestrial users.
This approach signifies a substantial improvement over the traditional IAB model, which relies on RF links for both backhaul and access \cite{Zhang2021f}.

Content caching refers to the strategic placement of frequently requested data at network edge nodes to reduce redundant transmissions and access latency, thereby optimizing network resource utilization \cite{Wang2014}. 
This approach has been extensively studied in terrestrial networks to improve the quality of experience (QoE) for users and reduce file delivery costs \cite{Qiu2002, Paschos2016, Li2018b}.
The concurrent distribution of identical content to multiple end-users introduces inherent spectral efficiency challenges in resource-constrained networks. 
While point-to-point unicast transmission requires redundant duplication of data streams for individual users, point-to-multipoint multicast transmission demonstrates significant efficiency gains by delivering common content to multiple subscribers through the same resource block~\cite{Vella2013, Araniti2017, Striccoli2019}.
However, traditional multicast, reliant on Steiner trees, often underutilizes available network paths, leading to suboptimal resource efficiency~\cite{Ho2008}.
The foundational work in \cite{Ahlswede2000} introduced {\em network coding}, a revolutionary technique that enables multicast transmissions to achieve the theoretical max-flow min-cut capacity, which was a bound previously unattainable with traditional routing approaches alone.
Network coding-based multicasting represents a paradigm shift that allows intermediate nodes to perform coding operations on packets from single or multiple sources rather than merely forwarding their received information packets \cite{Wang2006}.
The fundamental principle of network coding-based multicasting is that source data in a communication session can be algorithmically reconstructed by designated receivers via structured encoding, which eliminates the necessity for bitwise replication.
Building on this breakthrough, subsequent studies in \cite{Koetter2003, Li2003, Ho2006} established foundational principles for low-complexity code design for practical implementation of the network coding technique, which can be efficiently supported in HAP platforms through onboard processing units, such as specialized field-programmable gate array (FPGA)~\cite{Fragouli2006}. 
In this paper, we explore the integration of network coding-based multicasting with hybrid FSO/RF-enabled IAB in caching-capable multi-HAP networks. 
To the best of our knowledge, this work is the first to investigate the potential of this cost-effective design, which integrates network coding, content caching, and hybrid FSO/RF communications in multi-HAP networks to enhance rural connectivity. 
The contributions of this paper are elaborated in more detail later in this section.

\subsection{Related Studies}

Recent advancements in HAP networks have enabled more efficient caching mechanisms that can improve network performance across various applications and environments, as demonstrated in the literature. 
The authors in \cite{Yuan2023} proposed a convex optimization-based framework to jointly design the content caching and user assignment strategies to minimize transmission delay in a HAP-assisted network.
In \cite{Zhang2018c}, the use of HAPs for broadcasting cached content in vehicular networks to alleviate the traffic burden on roadside units (RSUs) was studied, and an analytical framework for the delay and hit-ratio of on-road mobile applications was developed. 
However, both \cite{Yuan2023} and \cite{Zhang2018c} assumed that all HAPs in the considered systems are directly connected to the core network, with no constraints on caching capacity and negligible transmission operating costs for communication between the core network and aerial HAPs. These assumptions may not hold in scenarios where HAPs are deployed to enhance network connectivity in rural and remote  areas.
Considering the limited caching capacity and backhaul transmission rate of aerial base stations (BSs), the authors in \cite{Kalantari2020} addressed the 3D placement of aerial nodes to minimize transmission power consumption. In \cite{Zhang2022d}, a stochastic geometry-based analytical framework was developed to analyze the outage probability in integrated satellite-HAP networks under a non-orthogonal multiple access (NOMA) scheme. The study in \cite{Vallero2022} investigated the performance of cache-enabled HAPs in urban networks to balance the load of terrestrial BSs.
While limited storage capacity was considered in \cite{Kalantari2020, Zhang2022d, Vallero2022}, these works neglected the links between aerial BSs and thus overlooked the potential benefits of multi-hop routing.

The integration of FSO as backhaul and RF as access links is a prevalent configuration in diverse HAP-aided communication networks \cite{Yahia2023, Huang2021b, Lee2022}.
In \cite{Yahia2023}, the authors derived closed-form expressions for the outage probability and bit error rate in HAP-assisted aerial-terrestrial downlink multicast transmissions.
By using HAPs as relays to assist the uplink transmission from ground users to the satellite, the authors in \cite{Huang2021b} proposed a space division multiple access (SDMA) scheme to maximize the overall ergodic rate.
In \cite{Lee2022}, the authors addressed the throughput maximization problem in multi-HAP-aided networks by optimizing the deployment positions of HAPs, their associations with ground BSs, and the transmit power allocation.
In \cite{Wang2024d}, a reconfigurable intelligent surface (RIS) is used to assist the uplink transmission of ground users to a UAV over RF access links. 
In that setup, the authors analyzed the secrecy performance and outage probability of the system under FSO-based backhaul links between the UAV and the HAP. 
In \cite{Liu2020a}, the authors studied the outage probability in a satellite-aerial-terrestrial communication system, where the aerial BS is equipped with multi-antenna technology to provide RF access links to ground users and an FSO receiver to receive signals from the satellite.
The authors in \cite{Ajam2020} considered employing  an aerial FSO-backhauled BS that buffers data received from ground users via RF access links and conducted an analysis of the ergodic sum rate.
In \cite{Che2021}, the authors presented a convex optimization algorithm to maximize system-level energy efficiency in an aerial-terrestrial communication system, where the aerial BS simultaneously harvested energy and decoded information from FSO backhaul links.
However, all the aforementioned studies do not take into account the ability to cache popular content at aerial BSs, and the joint access/backhaul design in multi-aerial-BS-assisted networks remains largely unexplored.

\subsection{Contribution}\label{sec:contribution}

This paper focuses on achieving cost-effective content delivery across wide underserved areas to bridge the digital divide.
We integrate caching techniques into multi-hop IAB HAP networks, utilizing FSO-based backhaul and RF-based access links. 
In the proposed caching-enabled HAP network, each HAP is equipped with a local cache of limited storage capacity and a multi-antenna RF array to serve ground users via beamformed directional transmissions.
Given the limited storage capacity of HAPs and the time-varying content retrieval requests from ground users, a proactive caching strategy is essential for enhancing QoE and reducing backhaul traffic. 
In addition, optimizing backhaul and access transmissions jointly, including beamforming design for spatially focused RF access links, is critical to reduce overall power consumption, while ensuring ground users' demands are met and the proactive caching strategy is effectively implemented.
We aim to address these challenges, and the main contributions are summarized as follows
\begin{itemize}
	\item 
	We propose a multi-HAP-based framework to provide content delivery services for underserved regions, leveraging network coding techniques for multicasting content over multi-hop FSO backhaul links. 
	Given system parameters such as network topology, content access requests,  and time-varying channel conditions, we formulate a long-term  weighted power cost minimization problem. 
	This problem jointly optimizes proactive caching strategies, beamforming vectors for directional RF transmissions, and network coding-based backhaul routing to ensure efficient resource utilization across both access and backhaul networks.
	\item 
	To solve the formulated problem, we propose a hierarchical algorithm incorporating both DRL and convex optimization.
	Specifically, the DRL agent is employed to execute the proactive caching strategy and decompose the multi-stage problem into successive deterministic resource allocation problems for each time slot. 
	\item 
	With the aid of numerical results, we prove the effectiveness of our proposed hierarchical solution by comparing its performance against several baseline algorithms.
	In addition, we draw a handful of design insights by investigating the effect of several system parameters, such as the FSO/RF channel bandwidth and data rate requirements.
\end{itemize}

The notations used throughout this paper are defined as follows. Bold uppercase letters represent matrices, while bold lowercase letters represent column vectors. 
The symbols $\mathbf{X}^{\rm H}$, $\rank(\mathbf{X})$, and $\Tr(\mathbf{X})$ correspond to the Hermitian (conjugate) transpose, the rank, and the trace of matrix ${\bf X}$, respectively. 
The notation ${\bf X}\succeq 0$ indicates that ${\bf X}$ is a positive semi-definite matrix. For any vector $\mathbf{x}$, $\mathbf{x}(i)$ refers to its $i$-th entry.

\begin{small}
	\begin{table}[ht]
		\caption{List of key notations.}
		\centering
		\begin{tabular}{c|c}
			\hline \hline
			Notation &  Description \\
			\hline
			$K, D, U$ & Numbers of HAPs, DCs, and UEs\\
			\hline
			$\mathcal{K}, \mathcal{D}, \mathcal{U}$ & Sets of HAPs, DCs, and UEs\\
			\hline
			$C$ & Number of content types in the network\\
			\hline
			$L$ & Maximum possible number of directed backhaul links\\
			\hline
			$M$ & Number of antennas at each HAP\\
			\hline
			$\mathbf{Z}^t$ & Binary matrix indicating cache placement at time $t$\\
			\hline
			$z_{k,c}^t$ & Binary indicator of content $c$ cached at HAP $k$ at time $t$\\
			\hline
			$N^{\mathrm{sto}}$ & Maximum number of contents cacheable at each HAP\\
			\hline
			$\mathcal{U}_k$ & Set of users served by HAP $k$\\
			\hline
			$\alpha_{u,c}^t$ & Binary indicator if user $u$ requests content $c$ at time $t$\\
			\hline
			$\beta_{k,c}^t$ & Data rate demand for content $c$ at HAP $k$ at time $t$\\
			\hline
			$\mu_c^{\rm cac}$ & Data rate required for caching content $c$\\
			\hline
			$\mu_c^{\rm acc}$ & Data rate required for accessing content $c$\\
			\hline
			$\digamma_{k,c}^{\mathrm{cac}} (t)$ & Caching request rate for content $c$ from HAP $k$ at time $t$\\
			\hline
			$\digamma_{k,c}^{\mathrm{acc}} (t)$ & Backhaul demand rate for content $c$ at HAP $k$ at time $t$\\
			\hline
			$\mathcal{S}_{c}(t)$ & Set of sources for content $c$ at time $t$\\
			\hline
			$\mathcal{K}^{\prime}_{c}(t)$ & Set of HAPs requesting content $c$ at time $t$\\
			\hline
			$\mathcal{K}^{\mathrm{cac}}_{c}(t)$ & Set of HAPs caching content $c$ at time $t$\\
			\hline
			$\mathcal{K}^{\mathrm{acc}}_{c}(t)$ & Set of HAPs accessing content $c$ at time $t$\\
			\hline
%			$e_{l,t}^{c,k}$ & Conceptual flow rate on backhaul link $l$ for session $c,k$\\
%			\hline
			$\eta_{l,t}^{c, {\rm cac}}$ & Actual flow rate for caching content $c$ on link $l$\\
			\hline
			$\eta_{l,t}^{c, {\rm acc}}$ & Actual flow rate for accessing content $c$ on link $l$\\
			\hline
			$\gamma_{l,t}$ & Total capacity of backhaul link $l$ at time $t$\\
			\hline
			$I(n), O(n)$ & Sets of incoming and outgoing links at node $n$\\
			\hline
			$h_{l,t}$ & Channel gain of FSO backhaul link $l$ at time $t$\\
			\hline
			$h^{\mathrm{al}}_{l}$ & Attenuation loss of FSO link $l$\\
			\hline
			$h^{\mathrm{at}}_{l,t}$ & Atmospheric turbulence loss of FSO link $l$ at time $t$\\
			\hline
			$h^{\mathrm{pl}}_{l,t}$ & Pointing error loss of FSO link $l$ at time $t$\\
			\hline
			$d_{l}$ & Distance of backhaul link $l$ in meters\\
			\hline
			$\nu$ & Visibility in kilometers for FSO links\\
			\hline
			$\lambda$ & Wavelength of optical signal in nanometers\\
			\hline
			$\kappa$ & Visibility coefficient in Kruse's model\\
			\hline
			$B_{\rm FSO}$ & Bandwidth of FSO links\\
			\hline
			$B_{\rm RF}$ & Bandwidth of RF links\\
			\hline
			$\tau_{l,t}$ & Time fraction allocated to FSO backhaul link $l$ at time $t$\\
			\hline
			$\rho_{l,t}$ & Achievable data rate of FSO link $l$ at time $t$\\
			\hline
			$p_{l,t}$ & Transmit power of FSO link $l$ at time $t$\\
			\hline
			$\sigma_{\rm FSO}^2$ & Variance of noise at optical receiver\\
			\hline
			$\mathcal{G}_{k,c}^t$ & Set of users requesting content $c$ from HAP $k$ at time $t$\\
			\hline
			$\mathbf{w}_{k,c,t}$ & Beamforming vector for HAP $k$, content $c$ at time $t$\\
			\hline
			$\mathbf{h}_{k,i,t}$ & Channel vector from HAP $k$ to user $i$ at time $t$\\
			\hline
			$\sigma_{\rm RF}^2$ & Power of AWGN on RF links\\
			\hline
			${\rm SINR}_{i}^{c}(t)$ & SINR of user $i$ for content $c$ at time $t$\\
			\hline
			$P_{\rm FSO}^{\rm HAP}(t)$ & FSO power consumption of HAPs at time $t$\\
			\hline
			$P_{\rm FSO}^{\rm DC}(t)$ & FSO power consumption of DCs at time $t$\\
			\hline
			$P_{\rm RF} (t)$ & RF power consumption of all HAPs at time $t$\\
			\hline
			$\omega$ & Weighting factor for HAP power consumption\\
			\hline
			$PC(t)$ & Total network power cost at time $t$\\
			\hline
			$p_{\max}$ & Maximum transmit power of FSO transmitter\\
			\hline
			$\delta_c$ & Target SINR for content $c$ over access links\\
			\hline \hline
		\end{tabular}
		\label{Tab:notation}
	\end{table}
\end{small}

\section{System Model}\label{sec:3}

\subsection{Network Model}\label{sec:net_mod}

\begin{figure*}[!htb] \centering \centering
	\begin{minipage}{0.45\linewidth}
		\centering
		\subfloat[]{\label{fig:system_model}\includegraphics[width=\linewidth]{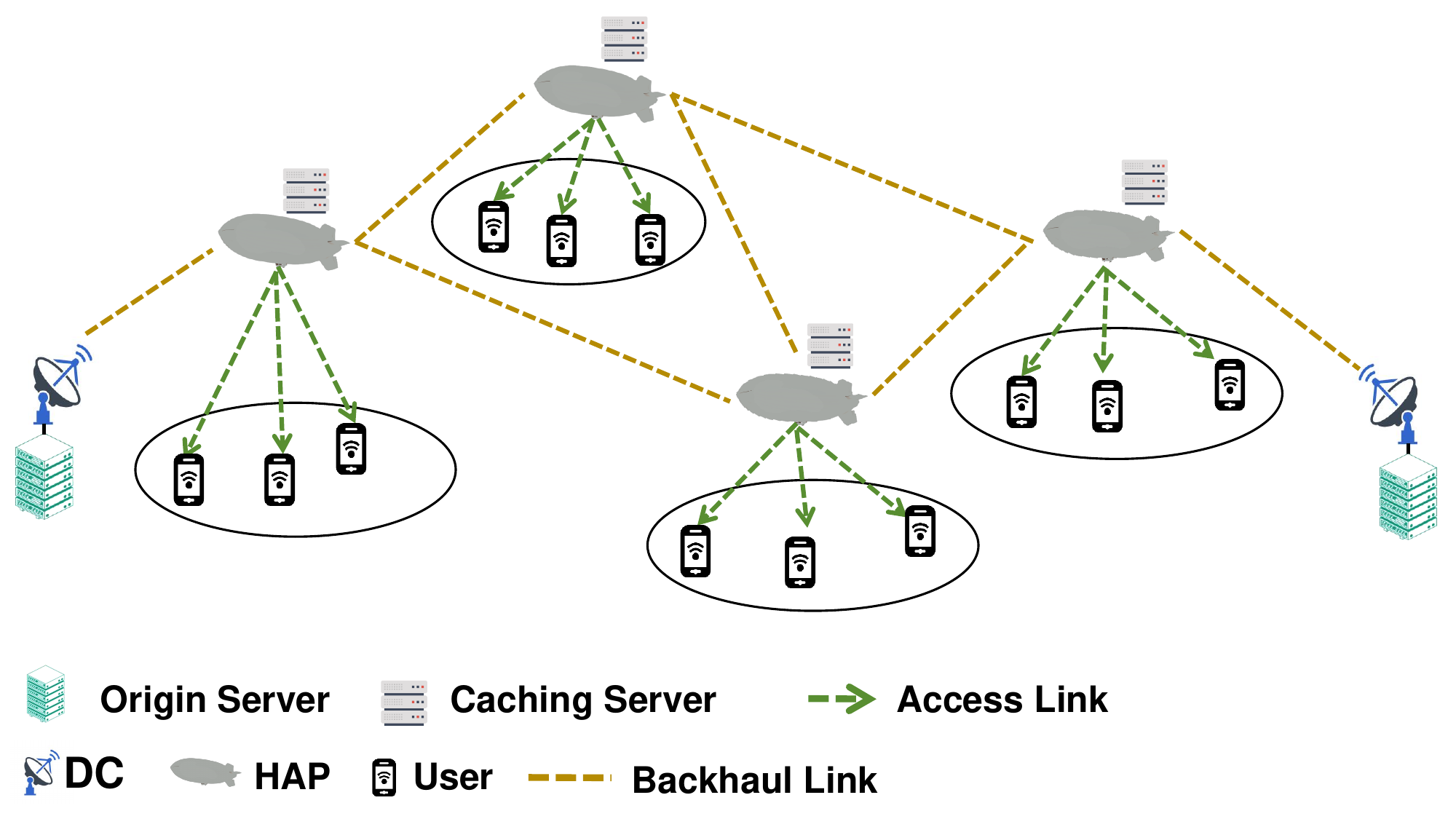}}
	\end{minipage}%
	\hspace{1mm}
	\begin{minipage}{0.45\linewidth}
		\centering
		\subfloat[]{\label{fig:system_workflow}\includegraphics[width=1\linewidth]{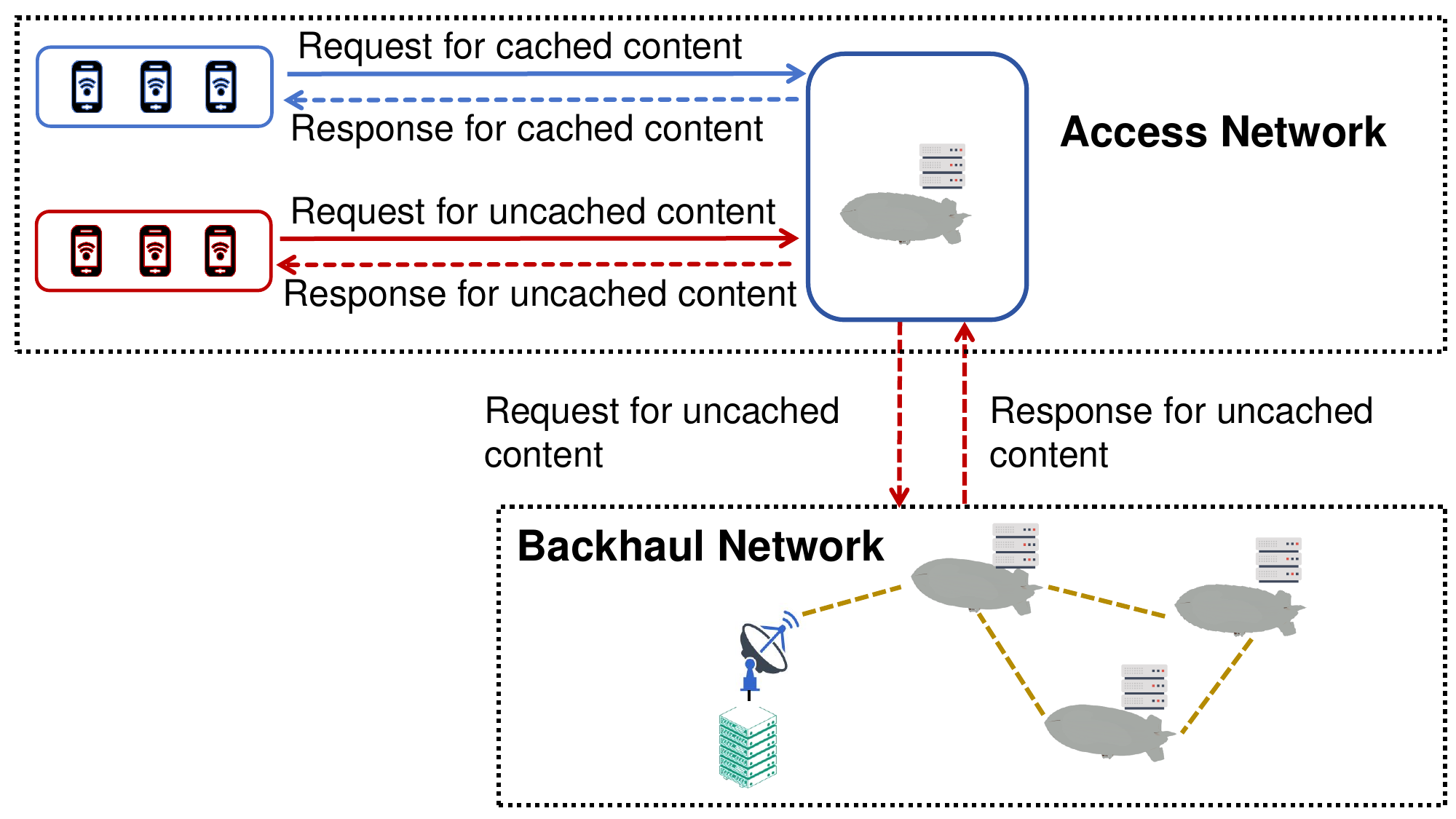}}
	\end{minipage}%
	\hspace{1mm}
	\caption{System model: (a) network architecture, (b) illustration of the workflow for servicing content requests from users.}
	\label{fig:sys}
\end{figure*}

We consider a network architecture comprising a set $\mathcal{K}$ of $K$ HAPs, a set $\mathcal{U}$ of $U$ ground users, and a set $\mathcal{D}$ of $D$ remote terrestrial data centers (DCs).
As shown in \figref{fig:system_model}, each HAP, equipped with an $M$-antenna RF transmitter and an FSO transceiver, offers RF access links to ground users and establishes FSO backhaul connections with other HAPs, respectively.
All the ground users are equipped with a single RF antenna to receive the signals from HAPs.
The ground data centers are cascaded with FSO transmitters and connected to the origin server.
It is assumed that each remote data center is exclusively associated with one HAP via a dedicated feeder link. 
Thus, the maximum possible number of the directed backhaul links in the network is $L = D + K(K - 1)$. 
In this paper, we consider a discrete-time system model where time is segmented into $T$ periods referred to as time slots, which is denoted by $t=1,\dots, T-1$.

\figref{fig:system_workflow} illustrates the workflow for processing content access requests from users.
If a HAP receives a request for content it has already cached, it fulfills the request directly via the access link. Conversely, if the content is not cached locally, the request is forwarded via backhaul links to other servers that contain the requested content type.
It is assumed that there are $C$ types of content in the network, all of which are available at DCs.
We define a ${K \times C}$ binary matrix $\mathbf{Z}^t$ to indicate the cache placement profile at HAPs in time slot $t$, where $z_{k,c}^t=1$ indicates that the $c$-th content is cached at the $k$-th HAP and $z_{k,c}^t=0$ otherwise. 
Each HAP is equipped with a local cache with limited storage capacity to store a subset of the content available across the network. 
Let $N^{\mathrm{sto}}$ be the maximum number of contents that can be cached at the HAP, the cache placement at the HAP $k$ is constrained by
\begin{align}\label{eq:CacConst}
	\sum\limits_{c=1}^C z_{k,c}^t \leq N^{\mathrm{sto}}.
\end{align}

% Caching request
To optimize system performance, HAPs must intelligently manage their limited storage by deleting cached content and fetching the desired content from data centers or other HAPs that have already cached it.
In this context, the caching request for content $c$ from HAP $k$ in time slot $t$ is expressed as:
\begin{align}\label{eq:ReqCacRate}
	\digamma_{k,c}^{\mathrm{cac}} (t) =
	\begin{cases}
		\mu_c^{\rm cac},	& \text{if}~z_{k,c}^{t+1}>z_{k,c}^{t}, \\
		0,	& \text{otherwise}, \\
	\end{cases}
\end{align}
where $\mu_c^{\rm cac}$ represents the data rate required for successfully caching content $c$ at HAPs.

% Access content demand
Let $\mathcal{U}_k$ denote the set of users served by the $k$-th HAP.
We assume that the sets $\mathcal{U}_k$ for $k\in\mathcal{K}$ are disjoint, i.e., $\cup_{k\in\mathcal{K}} \mathcal{U}_k=\mathcal{U}$, $ \mathcal{U}_k \cap \mathcal{U}_{\tilde{k}}= \emptyset$, $\forall k \neq \tilde{k}$.
We denote $\mu_{c}^{\rm acc}$ as the data rate requirement for users accessing content $c$.
The data rate demand for content $c$ at HAP $k$, resulting from serving its associated user set, is given by
\begin{align}\label{eq:AggReqSerRate}
	\beta_{k,c}^t  =  \mu_{c}^{\rm acc} \max\left( 0, \min \left(1, \sum\limits_{u\in \mathcal{U}_k}\alpha_{u,c}^t \right) \right) ,
\end{align} 
where $\alpha_{u,c}^t$ is an indicator variable that equals $1$ if user $u$ requests content 
$c$, and $0$ otherwise.
We assume that a typical user requests only one type of content at any given time slot, i.e., $ \sum\limits_{c=1}^C \alpha_{u,c}^t = 1,~\forall t, \forall u$.
Consequently, the backhaul demand for content $c$ at HAP $k$ from the content accessing requests from its covered users is 
\begin{align}\label{eq:ReqSerRate}
	\digamma_{k,c}^{\mathrm{acc}} (t)  =  (1-z_{k,c}^t) \beta_{k,c}^t.
\end{align} 

We apply network coding techniques to facilitate multicast sessions for transmitting content over backhaul links.
For instance, the process of transmitting content $c$ from the set of sources $\mathcal{S}_{c}(t) = \{ k \vert z_{k,c}^t > 0, k \in \mathcal{K}\} \cup \mathcal{D}$ to the set of HAPs $\mathcal{K}^{\prime}_{c}(t) = \{ k \vert \digamma_{k,c}^{\mathrm{cac}} (t) > 0 \text{ or } \digamma_{k,c}^{\mathrm{acc}} (t) > 0 , k \in \mathcal{K}\}$ can be viewed as a multicast session.
Additionally, due to the differing data rate requirements for content caching and content accessing requests, the multicast session with respect to content $c$ is a bi-rate multicast session, which includes
\begin{enumerate}
	\item A content caching session to the set of destinations $\mathcal{K}^{\mathrm{cac}}_{c}(t) = \{ k \vert \max\big(\digamma_{k,c}^{\mathrm{cac}} (t) ,\digamma_{k,c}^{\mathrm{acc}} (t)\big)=\mu_{c}^{\rm cac}, k \in \mathcal{K}\}$ with the data rate requirement $\mu_{c}^{\rm cac}$, 
	\item A content accessing session to the set of destinations $\mathcal{K}^{\mathrm{acc}}_{c}(t) = \{ k \vert \max\big(\digamma_{k,c}^{\mathrm{cac}} (t) ,\digamma_{k,c}^{\mathrm{acc}} (t)\big)=\mu_{c}^{\rm acc}, k \in \mathcal{K}\}$  with the data rate requirement $\mu_{c}^{\rm acc}$.
\end{enumerate}

The fundamental study in \cite{Ahlswede2000} demonstrates that multicast throughput with network coding is achievable if the throughput from the source to each destination, considered independently as a unicast, is feasible.
For each backhaul link $l$, we denote $e_{l,t}^{c,k^{\rm cac}}$ and $e_{l,t}^{c,k^{\rm acc}}$ as the conceptual flow rate on backhaul link $l$ in the multicast sessions of caching content $c$ at at the destination HAP $k^{\rm cac}$ and for fulfilling access requests for content $c$ from the HAP $k^{\rm acc}$ to its serving users, respectively.
The actual flow rates for content caching and accessing sessions on backhaul link $l$ are denoted by 
 $\eta_{l,t}^{c, {\rm cac}}$ and $\eta_{l,t}^{c, {\rm acc}}$, respectively.
The routing constraints for the multi-hop FSO-based backhaul network are defined as follows \cite{Lakshminarayana2013} % Yuan2006
\begin{align}
	\sum\limits_{l \in I(k^{\rm cac} )} 
	\hspace{-0.25em}  &
	e_{l,t}^{c, k^{\rm cac}  } 
	\geq \mu_c^{\rm cac},  ~ 
	\forall k^{\rm cac} \in 
	\mathcal{K}^{\rm cac}_{c}(t), \forall c,\forall t,  \label{eq:NCQosCac} \\
			\sum\limits_{l \in I(k^{\rm acc} )} \hspace{-0.25em}  &
			e_{l,t}^{c,k^{\rm acc} } 
			 \geq \mu_c^{\rm acc},  ~ \forall k^{\rm acc} \in 
			\mathcal{K}^{\rm acc}_{c}(t), \forall c,\forall t, 
			\label{eq:NCQosAcc} \\
	\sum\limits_{l \in I(n)} \hspace{-0.25em}
	& e_{l,t}^{c, k'}  = 
	\sum\limits_{l' \in O(n)} 
	\hspace{-0.25em} 
	e_{l',t}^{c,k'}, ~ \forall n \in \mathcal{K},  n \neq k',n  \not\in 
	\mathcal{S}_c(t), \notag\\
	& \qquad \qquad \qquad \forall k' \in \mathcal{K}^{\rm cac}_{c}(t) 
	\cup \mathcal{K}^{\rm acc}_{c}(t), \forall c, \forall t,
	\label{eq:clowConservation} \\
			\sum\limits_{n \in \mathcal{S}_c(t)} &
			\sum\limits_{l \in I(n)} \hspace{-0.25em}
			 e_{l,t}^{c, k^{\rm cac}} \geq \mu_c^{\rm cac}, ~
			 \forall k^{\rm cac}\in \mathcal{K}^{\rm cac}_{c}(t), \forall c, \forall t,
			\label{eq:origin_cflowCac} \\
			\sum\limits_{n \in \mathcal{S}_c(t)} &
			\sum\limits_{l \in I(n)} \hspace{-0.25em}
			e_{l,t}^{c, k^{\rm cac}} \geq \mu_c^{\rm acc}, ~
			\forall k^{\rm acc}\in \mathcal{K}^{\rm acc}_{c}(t), \forall c, \forall t,
			\label{eq:origin_cflowAcc}  \\
	e_{l,t}^{c,k^{\rm cac}}&  \leq  \eta_{l,t}^{c, {\rm cac}},  ~ 
	\forall l, \forall k^{\rm cac} \in \mathcal{K}^{\rm cac}_{c},
	\forall c, \forall t,  \label{eq:cflowReqCac} \\
			e_{l, t}^{c,k^{\rm acc}} &  \leq  \eta_{l,t}^{c, {\rm acc}},  ~ 
			\forall l, \forall k^{\rm acc} \in \mathcal{K}^{\rm acc}_{c}(t),
			\forall c, \forall t, \label{eq:cflowReqAcc} \\
			\sum\limits_{c=1}^C  
			\big(  & \eta_{l,t}^{c, {\rm cac}}  + \eta_{l,t}^{c, {\rm acc}} 
			\big) \leq  \gamma_{l,t},  ~ \forall l, \forall t,  
			\label{eq:pflowCap} 
\end{align}
where $ I(\cdot)$ denotes the set of links incoming to a node,  $O(\cdot)$  denotes the set of links outgoing from it.
The QoS constraints \eqref{eq:NCQosCac} and \eqref{eq:NCQosAcc} ensure that the multicast rates for caching and accessing content $c$ at any destination meet or exceed the required data rates.
The conceptual flow conservation constraint \eqref{eq:clowConservation} mandates that for any intermediate node $n$, the sum of incoming flow rates for content $c$ equals the sum of outgoing flow rates, thereby ensuring no data loss within the network.
Constraints \eqref{eq:origin_cflowCac} and \eqref{eq:origin_cflowAcc} guarantee that the total sum of conceptual flow rates originating from the set of source nodes $ \mathcal{S}_c(t)$  for content caching and accessing meets or exceeds the corresponding data rate requirements.
Constraints \eqref{eq:cflowReqCac} and \eqref{eq:cflowReqAcc} specify that the actual flow rate of each multicast sub-session on backhaul link $l$ is determined by the maximum, rather than the sum, of the conceptual flow rates on that link to all destinations, which is an advantage of using network coding. 
Additionally, constraint \eqref{eq:pflowCap} ensures that the total physical flow rate on link \(l\) at time \(t\) does not exceed its capacity \(\gamma_{l,t}\).

\subsection{Communication Model} 

The channel gain of the FSO backhaul link $l$ is given by
\begin{align}\label{eq:FsoModel}
	h_{l,t} = h^{\mathrm{al}}_{l} h^{\mathrm{at}}_{l,t} h^{\mathrm{pl}}_{l,t},
\end{align}
where $ h^{\mathrm{al}}_l $ represents the attenuation loss due to the absorption and scattering effects in the atmosphere, $ h^{\mathrm{at}}_{l,t} $ accounts for the atmospheric turbulence loss resulting from the scintillation of
the FSO signal, and $h^{\mathrm{pl}}_{l,t}$ is the pointing error loss induced by stochastic fluctuations of the transmitted beam in the angle-of-arrival at the receiver aperture.

The attenuation loss component of the FSO link $l$ can be modeled as follows
\begin{align}\label{eq:FsoAttLoss}
	h^{\mathrm{al}}_{l} = \exp\left(
	\frac{-0.0009
	}
	{\nu} 
	\left( \frac{\lambda}{550}\right)^{\kappa} d_{l} 
	\right),
\end{align}
where $d_{l}$ is the distance in meters of the backhaul link $l$,  $\nu$ is the visibility in kilometers, and $\lambda$ is the wavelength of the optical signal in nanometers.
The coefficient $\kappa$ is determined by Kruse's model, which accounts for different visibility conditions as follows \cite{Kruse1962}
\begin{align}\label{eq:visbility}
	\kappa =
	\begin{cases}
		1.6,	& \nu > 50~{\rm km}, \\
		1.3,	&   6~{\rm km}\leq \nu \leq 50~{\rm km}, \\
		0.585 \nu^{1/3}, & \nu  < 6~{\rm km}.
	\end{cases}
\end{align}

For tractability, we assume a fixed visibility parameter $\nu$ in \eqref{eq:visbility}. 
Incorporating time-varying visibility and its impact on FSO attenuation requires location-specific weather models and is left for future works.
Compared to the lognormal and Gamma-Gamma distributions, the exponentiated Weibull distribution has demonstrated superior accuracy in fitting experimental data under weak and moderate turbulence conditions~\cite{Barrios2012}.
Motivated by this, in this paper, the atmospheric turbulence loss on the FSO link $l$ is modeled using the exponentiated Weibull distribution, which is given by
\begin{align}\label{eq:FsoAtmTurb}
	& \FuncSty{f}_{h^{\mathrm{at}}_{l,t}}(h) = \notag\\
	&  	\quad 
	\frac{\varphi \varsigma}{\varepsilon} \left(\frac{x}{\varepsilon}\right)^{\varsigma-1} \exp \left[-\left(\frac{h}{\varepsilon}\right)^{\varsigma}\right] \left\{1 - \exp\left[-\left(\frac{h}{\varepsilon}\right)^{\varsigma}\right]\right\}^{\varphi-1},
\end{align}
where $\varphi$ and $\varsigma$ are shape parameters influenced by the receiver aperture diameter and the scintillation index, and $\varepsilon$ is the scale parameter.

According to \cite{Bashir2023}, the probability distribution function of the pointing error loss on the FSO link $l$ can be expressed as 
\begin{align}
	\FuncSty{f}_{h^{\mathrm{pl}}_{l,t}} (h) 
	= \frac{\vartheta^2}{\sigma_0^2} 
	\left( 
	\frac{2 \vartheta^2 \Upsilon_l^2}{\chi_{\rm d}^2}
	\right)
	^{ \frac{\vartheta^2}{\sigma_0^2} } 
	h^{(\vartheta-\sigma_0^2) / \sigma_0^2} 
	\mathbbm{1}_{[0, \frac{\chi_{\rm d}^2}{2 \vartheta^2 \Upsilon_l^2})} (h),
\end{align} 
where $\chi_{\rm d}$ is the radius of the optical receiver aperture, $\sigma_0$   is the standard deviation of the angular pointing error, $\vartheta$ is the angular beamwidth, and $\Upsilon_l$ is the transmission distance of link $l$.

Using transmit power  $p_l$, the achievable data rate of an FSO link under intensity modulation direct detection (IM/DD) can be expressed as \cite{Huang2019}
\begin{align}\label{eq:FsoRate}
	\rho_{l,t} = \frac{B_{\rm FSO}}{2\ln 2}
				\ln\left(
						1+\frac{\exp(1) \varrho^2 h_{l,t}^2 p_{l,t}^2 }
						{2\pi\sigma_{\rm FSO}^2}
						\right),
\end{align}
where $\varrho$ is the responsivity of the photodetector at the HAP, $B_{\rm FSO}$ is the bandwidth of the FSO link $l$, $\tau_l$ is the time-fraction allocated to link $l$, and $\sigma_{\rm FSO}^2$ is the variance of the zero-mean Gaussian noise at the optical receiver.

To avoid transmission conflicts, the sum of the fractions allocated to different backhaul links occurring at the same node should be no larger than $1$.
Let $\tau_{l,t}$ denote the time fraction allocated to FSO backhaul link $l$, we have 
\begin{align}
	& 	\sum\limits_{l\in I(k)} \tau_{l,t} + \sum\limits_{l^\prime \in O(k)} \tau_{l^\prime,t} \leq 1,\quad \forall k \in \mathcal{K} \label{eq:TimeFracHAP} \\
	& \sum\limits_{l^\prime \in O(d)} \tau_{l^\prime,t}  \leq 1,\quad\forall d \in \mathcal{D} \label{eq:TimeFracDC}.
\end{align}

Taking into account the time fraction, the minimum transmit power required in time slot $t$ to achieve a data rate of $\gamma_{l,t} = \tau_{l,t} \rho_{l,t}$ is given by
\begin{align}\label{eq:powerBHlink}
	p_{l,t} =  \sqrt{\frac{ \tau_{l,t}^2}{g_{l,t}} 
		\left(\exp\left(\frac{ 2\ln2 \gamma_{l,t}}{B_{\rm FSO}\tau_{l,t}} \right) -1 \right)},
\end{align}
where $g_{l,t} = \frac{\exp(1) \rho^2 h_{l, t}^2}{2\pi\sigma_{\rm FSO}^2}$.

The power consumption of HAPs and DCs for the FSO-based backhaul transmission in time slot $t$ can be as follows, respectively
\begin{align}
	P_{\rm FSO}^{\rm HAP}(t) = 	\sum\limits_{k\in \mathcal{K}} \sum\limits_{l\in O(k)} p_{l,t},
\end{align}
and 
\begin{align}
	P_{\rm FSO}^{\rm DC}(t)  = 	\sum\limits_{d\in \mathcal{D}} \sum\limits_{l\in O(d)} p_{l,t}.
\end{align}

Users associated with the $k$-th HAP can be categorized into $G_k$ groups based on the type of content they request.
Let $\mathcal{G}_{k,c}^t$ denote the set of users associated with HAP $k$ who request content $c$ during time slot $t$. 
The weights applied to the transmit-antenna elements of the $k$-th HAP for the $c$-th multicast group during time slot $t$ are represented by the beamforming vector $\mathbf{w}_{k,c,t} \in \mathbb{C}^M$.
The {\rm SINR} of the user $i \in \mathcal{G}_{k,c}^t$ is given by 
\begin{align}
	{\rm SINR}_{i}^{c}(t) = 
	\frac{|\mathbf{w}_{k,c,t}^{\rm H}\mathbf{h}_{k,i,t}|^2}
	{\sum\limits_{\tilde{c}=1,l\neq c}^C |
		\mathbf{w}_{k,\tilde{c},t}^{\rm H}\mathbf{h}_{k,i,t}|^2 
		+\sigma_{\rm RF}^2
	},
\end{align}
where $\mathbf{h}_{k,i,t} $ is an $M\times1$ complex vector that models the propagation loss and phase shift of the frequency-flat quasi-static channel from each RF transmit antenna array of HAP $k$ to the RF receive antenna of user $i$, and $\sigma_{\rm RF}^2$ represents the power of the additive white Gaussian noise (AWGN) on the RF links.

The total power output from the transmitting antenna arrays of all HAPs in time slot $t$ is calculated as follows
\begin{align}
	P_{\rm RF} (t) = \sum\limits_{c=1}^C \sum\limits_{k\in \mathcal{K}} 
							\|\mathbf{w}_{k,c,t}\|_2^2  .
\end{align}

\subsection{Problem Formulation}

The total network power cost for the considered architecture comprises two primary components: the power costs associated with the DCs and HAPs.
The power cost of DCs refers to the energy consumption from backhaul FSO transmission, while the transmission power cost of HAPs includes both FSO and RF transmissions.
Since HAPs are aerial platforms powered by limited on-board energy sources, we introduce a weighting factor $\omega$ to emphasize their contribution to the overall network power cost.
The total network power cost at time slot $t$ can be expressed as the following weighted sum
\begin{align}
	PC(t) = P_{\rm FSO}^{\rm DC}(t) + \omega \left( P_{\rm FSO}^{\rm HAP}(t) + P_{\rm RF} (t)\right).
\end{align}

Our objective is to derive the optimal joint proactive caching strategy, backhaul network coding-enabled routing policy, and access beamforming design to minimize the weighted sum of power costs by solving the following optimization problem
\begin{subequations}\label{eq:opt1}
	\begin{align}
		(\textbf{OPT-1}):~&\underset{\{\mathbf{Z}_t, \mathcal{E}_t, \mathcal{T}_t, \mathcal{W}_t \}_{0\leq t\leq T-1}}{\text {minimize}}~\, 
		\sum\limits_{t=0}^{T-1}   PC(t)  \label{eq:obj_fun1}\\
		\text {s.t.} \qquad & \qquad  
		\eqref{eq:CacConst},
		~ \eqref{eq:NCQosCac} - \eqref{eq:cflowReqAcc},
						~\eqref{eq:TimeFracHAP} -\eqref{eq:powerBHlink}, \notag\\
	 	&\hspace{-3.85em}	\sum\limits_{c=1}^C 
	 			\big(\eta_{l,t}^{c, {\rm cac}}  + \eta_{l,t}^{c, {\rm acc}}\big) 
	 	   \leq \frac{B_{\rm FSO}\tau_{l,t} }{2\ln 2} \ln(1+g_{l,t} p_{\max}^2 ),~\forall l,\\
%		& p_{l,t} \leq p_{\max}\\
		& {\rm SINR}_{i}^{c}(t) \geq \delta_c, ~ \forall i \in \mathcal{G}_{k,c}^t, \forall k \in\mathcal{K}, \forall 1 \leq c \leq C,
		& \label{eq:consBF}  \\
		& \mathbf{Z}_t \in \{0,1\}^{K \times C},
	\end{align}
\end{subequations}
where $\mathcal{E}_t = \{e_{l,t}^{c, k'} | \forall l, \forall c, \forall k' \in \mathcal{K}^{\rm cac}_{c}(t) 
\cup \mathcal{K}^{\rm acc}_{c}(t)\}$ represents the conceptual flow routing policy, $\mathcal{T}_t = \{\tau_{l,t} | \forall l\}$ is the time fraction allocation policy, $ \mathcal{W}_t=\{\mathbf{w}_{k,c,t}|\forall k, \forall c\}$ denotes the  beamforming policy, $p_{\max}$ indicates the maximum transmit power of the FSO transmitter, and $\delta_c$ is the target {\rm SINR} to achieve the transmission rate $\mu_c^{\rm acc} = B_{\rm RF}\log_2(1+\delta_c)$. 

\section{Optimization Framework}

\subsection{Overview of Proposed Solution}
It is challenging to solve problem (\textbf{OPT-1}) by using conventional numerical optimization-based methods, since it is a multi-stage mixed-integer non-convex programming problem. 
To address this issue, we propose a multi-level optimization scheme that integrates DRL and convex optimization techniques. 
In specific, we employ a DRL agent to manage the time-varying cache placement decisions of HAPs. 
The DNN component of the DRL agent not only focuses on immediate returns but also accounts for longer-term performance. 
Once the DRL agent specifies a caching decision for a typical time slot, the remaining resource allocation variables can be efficiently optimized using convex optimization-based approaches.
In essence, this framework leverages the DRL agent to guide caching decisions over time, while relying on efficient convex solvers to tackle the subproblems associated with each time slot once the caching decisions have been made.  
This hybrid architecture, which separates content caching decisions from resource allocation, significantly reduces computational complexity compared to pure end-to-end DRL approaches that would otherwise need to explore an intractably large combined action space.
In the following subsections, we introduce the proposed DRL-based caching mechanism and the convex optimization-based algorithms in detail.

\subsection{DRL-Based Caching Control Strategy}\label{sec:3b}
We adopt a Proximal Policy Optimization (PPO) agent to make the content cache placement decisions over the considered time slots. 
The agent decides the content cache placement profile on the HAPs for the subsequent time slot based on its observed network state.
In our context, the state of time slot $t$ is denoted by $\mathbf{s}_t$ and is constructed by the content placement on local caches of HAPs, and the content accessing requests indicators of users, i.e., $\mathbf{s}^t= [z_{1,1}^t,\dots, z_{K,C}^t, \alpha_{1,1}^t, \dots,\alpha_{U,C}^t ]$. 
This construction of the state space ensures the agent learns caching decisions based solely on observed requests without relying on a priori knowledge of content popularity distributions.

The generation of action in a PPO agent is performed by sampling from the probability distribution characterized by the output of its actor network $\pi_{\theta}$, which is a neural network with parameter set ${\theta}$. 
To cope with the multi-binary action space requirement of the formulated optimization problem, we build up an actor network to output the parameter vector for an independent $KC$-dimensional multivariate Bernoulli distribution, denoted by $\pi_{\theta}(\mathbf{a}_t|\mathbf{s}_t)$, based on the input state $\mathbf{s}_t$.
For a typical action $\mathbf{a}_t \in \{0,1\}^{KC}$, we propose a function \FuncSty{ProcAct}, as detailed in Algorithm \ref{alg:ProcAct}, to ensure compliance with the local cache storage capacity constraint given in \eqref{eq:CacConst}.
As described in line 6 of Algorithm \ref{alg:ProcAct}, if the total number of cached contents at a typical HAP specified by the action $\mathbf{a}^t$ exceeds $B^{\mathrm{sto}}$, this HAP can only store the first $B^{\mathrm{sto}}$ types of desired contents.

\begin{algorithm}[hbt!]
	\small
	\SetAlgoLined
	\DontPrintSemicolon
	\KwIn{$\mathbf{a}_t $ }
	\KwOut{$\mathbf{Z}^{t+1}$ }
	\SetKwFunction{FMain}{ProcAct}
	\SetKwProg{Fn}{Function}{:}{}
	\Fn{\FMain{$\mathbf{a}^t$}}
	{\For{$k\gets1$ \KwTo $K$}{
			\nosemic $i \leftarrow 0$ \;
			\For{$c \leftarrow 1$}{
				\nosemic $j \leftarrow  (k-1) C + c $ \;
				\uIf{ $\mathbf{a}_t(j) = 1$ \textbf{and} $i  < B^{\mathrm{sto}} $ }{
					\nosemic $z_{k,c}^{t+1} \leftarrow 1$  \;
					\nosemic $i\leftarrow i + 1$  \;} 
				\uElse{ \nosemic $z_{k,c}^{t+1} \leftarrow 0$ }}}  
		\Return $ \mathbf{Z}^{t+1}$}
	\caption{Processing action.}
	\label{alg:ProcAct}
\end{algorithm}

In addition to the actor network, PPO employs a critic network $V_{\phi}$, parameterized by $\phi$, to estimate the expected long-term returns from a given state.
This estimation aids the actor network in refining its policy by assessing the quality of the actions taken.
Given the action $\mathbf{a}_t$ generated from the parameterized policy $\pi_{\theta}$,  the instantaneous power cost in time slot $t$ can be determined by using the algorithms presented in the following subsections, which solve the transmitting power cost minimization problem as follows:
\begin{align}
	q_t^* := \underset{ \{\mathcal{E}(t), \mathcal{T}(t), \mathcal{W}(t)\} }{\min } PC(t).
\end{align}

The objective of PPO is to derive the optimal stochastic policy $\pi_{\theta*}$ for content cache placement control that maximizes the weighted cumulative reward, thereby minimizing the long-term cumulative power cost:
\begin{align}\label{eq:goal_arl}
	\pi_{\theta^*} := \underset{\pi_{\theta}}{\arg \max} 
	~
	{\mathbbm{E}} \left[ \sum\limits_{t=0}^{T-1} \gamma^{t} r_t\right],
\end{align}
where $r_t=-q_t^* $ and $\gamma$ denote the instantaneous reward in time slot $t$ and the discount factor for future rewards, respectively.

After collecting environmental interaction samples in the format $\{( {\bf s}_t, {\bf a}_t, r_t, {\bf s}_{t+1})\}$, as explicitly detailed in \cite{ppo2017},  the learning iteration of the actor network in PPO is performed using mini-batch stochastic gradient descent (SGD) to optimize the following clipped surrogate loss function:
\begin{align}\label{eq:update_theta}
	\mathrm{Loss}({\theta}) & =   \frac{1}{N_B} \sum\limits_{n=0}^{N_B-1}  
	\Biggl[
	\min  \Biggl(
	\frac{\pi_\theta(\mathbf{a}_n|\mathbf{s}_n)}{\pi_{\theta^\prime} (\mathbf{a}_n|\mathbf{s}_n)}                     
	\hat{A}_{\pi_{\theta^\prime}}\left(\mathbf{s}_n,\mathbf{a}_n \right), \notag\\
	& \FuncSty{clip}
		\left(\frac{\pi_\theta(\mathbf{a}_n|s_n)}{\pi_{\theta^\prime} (\mathbf{a}_n|\mathbf{s}_n)}, \left[1- \epsilon, 
			1+\epsilon\right]\right) 
		\hat{A}_{\pi_{\theta^\prime}}\left(\mathbf{s}_n,\mathbf{a}_n\right)
	\Biggr)
	\Biggr],
\end{align}
where $N_B$ is the mini-batch size, $\epsilon$ is the clipping ratio that guarantees $\theta$ cannot dramatically deviate from $\theta^\prime$, and $\hat{A}_{\pi_{\theta}}$ denotes the estimation of the advantage for a state-action pair. 
$\FuncSty{clip}\left(x, \left[x_{\min},x_{\max} \right]\right)$ returns $x$ for $x_{\min}\leq x \leq x_{\max} $, $x_{\min} $ for $x<x_{\min}$, and $x_{\max}$ for $x>x_{\max}$. 
This \FuncSty{clip} operation in \eqref{eq:update_theta} is designed to prevent the new policy from deviating substantially from the old policy $\pi_{\theta^\prime}$ by capping the probability ratio at $1 + \epsilon$ for positive advantage estimates and at $1 - \epsilon$ for negative advantage estimates.

The $\hat{A}_{\pi_{\theta}}(s_t,a_t)$ can be estimated by the generalized advantage estimator (GAE)~\cite{schulman2015high}, which is given by
\begin{align}\label{eq:gae}
	\hat{A}_{\pi_{\theta}}(\mathbf{s}_n,\mathbf{a}_n) 
	= 
	\sum\limits_{i=n}^{N_B-1} \left( \gamma \zeta \right)^{i-n} 
	\left(
	r_i+ \gamma {V}_{\phi}\left(\mathbf{s}_{i+1}\right) 
	- {V}_{\phi}\left(\mathbf{s}_{i}\right),
	\right)  
\end{align}
where $ \zeta \in (0,1)$ is the trace-decay parameter that serves to reduce variance and enhance training stability.

The value network $V_{\phi}$  is updated by minimizing the temporal difference between the predicted value $V_{\phi}(\mathbf{s}_n)$ and the target value $\hat{V}(\mathbf{s}_n)$ as follows
\begin{align}\label{eq:update_phi}
	\mathrm{Loss}({\phi}) = 
		\frac{1}{N_B}\sum\limits_{n=0}^{N_B-1} 
			\left(
				{V}_{\phi}({\bf s}
				_n) - \hat{V}({\bf s}_n)
			\right)^2,
\end{align}
in which 
\begin{align}\label{eq:cal_tar}
	\hat{V}(\mathbf{s}_n) = \sum\limits_{i=n}^{N_B-1} \gamma^{i-n}(r_{i-n}).
\end{align}

\subsection{Power Cost Minimization in Single Time Slot}
Once the content placement decision is determined by the DRL agent described above, the resource allocation for each time slot can be addressed using the algorithms presented in the following.
Additionally, the resource allocation for each time slot can be split into two subproblems: the first corresponding to the FSO based backhaul links, and the second focused on the beamforming design for RF-based access links.
For the purpose of tractability, we approximate the FSO channel capacity in \eqref{eq:FsoRate} as 
\begin{align}\label{eq:FsoRateApprox} 
	\rho_{l,t} \approx \frac{B_{\rm FSO} }{2\ln 2} \ln(g_{l,t}p_{l,t}^2),
\end{align}
which is a tight approximation for high SNR.
Indeed, a high SNR can be ensured since a minimum required system data rate is set to guarantee a desired backhaul transmission data rate.

Define the function $f_{l,t}(x)= \frac{1}{\sqrt{g_{l,t}}} \exp\left(\frac{ x \ln2 }{B_{\rm FSO}} \right)$.
It follows from \eqref{eq:FsoRateApprox} that the energy consumption on the FSO link $l$ in time slot $t$ can be expressed as  
\begin{align}\label{eq:PwrApprox} 
	\tilde{p}_{l,t}  =
	\begin{cases}
		\tau_{l,t}f_{l,t}( \frac{\gamma_{l,t}}{\tau_{l,t}}),	& \tau_{l,t} >0, \\
		0,	&  \tau_{l,t} = 0 \text{~or~} \gamma_{l,t}=0.
	\end{cases}
\end{align}

\begin{lemma}[] \label{lma:Persp}
	$\tilde{p}_{l,t}$ is jointly convex in $\gamma_{l,t}$ and $\tau_{l,t}$.
\end{lemma}
\begin{IEEEproof}
Since $f_{l,t}(x)$ is a convex function, its perspective function, i.e., $\tau_{l,t}f_{l,t}( \frac{\gamma_{l,t}}{\tau_{l,t}})$ , is also convex \cite{Moehle2015}. 
For the cases where $\tau_{l,t} = 0$ or $\gamma_{l,t} = 0$, the convexity of $\tilde{p}_{l,t}$ with respect to $\gamma_{l,t}$ and $\tau_{l,t}$ can be proved by employing the approach outlined in \cite{Wang2008} and hence omitted here.
\end{IEEEproof}

The objective of backhaul routing optimization is to obtain the optimal time fraction allocation policy $\mathcal{T}^*_t$ and data rate routing policy $\mathcal{E}^*_t$, in order to minimize the weighted sum of power consumed by DCs and HAPs.
To this end, based on \eqref{eq:PwrApprox}, minimizing the backhaul power cost can be achieved by solving the following reformulated problem
\begin{subequations}\label{eq:optNC}
	\begin{align}
		(\textbf{OPT-2})\!:\quad&\underset{ \{\mathcal{T}_t, \mathcal{E}_t \}}{\text {minimize}}~\, \sum\limits_{d\in \mathcal{D}} \sum\limits_{l\in O(d)}\tilde{p}_{l,t} + \omega  \sum\limits_{k\in \mathcal{K}} \sum\limits_{l\in O(k)} \tilde{p}_{l,t}  	 \label{eq:ObjFunNC}\\
		\text {s.t.} \qquad \qquad & \eqref{eq:NCQosCac} - \eqref{eq:pflowCap},
		~\eqref{eq:TimeFracHAP} -\eqref{eq:TimeFracDC} \notag\\
		&  \tilde{p}_{l,t} \leq \tau_{l,t} p_{\max}, ~\forall l. \label{eq:lastConstrOpt2}
	\end{align}
\end{subequations}

\begin{lemma}[] \label{lma:ConvexProf}
	The transformed optimization problem (\textbf{OPT-2}) is a convex optimization problem.
\end{lemma}
\begin{IEEEproof}
	It can be readily demonstrated that the constraints  \eqref{eq:NCQosCac} - \eqref{eq:cflowReqAcc} are convex, as each inequality is a linear combination of variables. Furthermore, the constraints  \eqref{eq:TimeFracHAP}-\eqref{eq:TimeFracDC} define a convex feasible set.
	The right-hand side of constraint \eqref{eq:lastConstrOpt2} is linear, implying that it is both convex and concave, while the left-hand side is convex due to the convexity of $\tilde{p}_{l,t}$ with respect to $\gamma_{l,t}$ and $\tau_{l,t}$.
	Consequently, constraint \eqref{eq:lastConstrOpt2} is convex.
	Since the convexity of the objective function has been established in Lemma \ref{lma:Persp}, it follows that the transformed problem (\textbf{OPT-2}) is a convex optimization problem.
\end{IEEEproof}

Lemma \ref{lma:ConvexProf} implies that problem (\textbf{OPT-2}) satisfies Slater's condition and has a zero-duality gap. 
This indicates that standard convex optimization tools, such as CVXPY \cite{Diamond2016} and MOSEK \cite{ApS2019}, can be effectively applied to solve this problem by using the primal-dual method, thereby yielding the optimal solutions for data rate and time fraction allocation.

Let $\mathbf{W}_{k,c,t}$ represent the beamforming matrix for HAP $k$ multicasting content $c$, defined as $\mathbf{W}_{k,c,t} =\mathbf{w}_{k,c,t} \mathbf{w}_{k,c,t}^H$.
The beamforming subproblem, with the objective of minimizing the power cost of the transmission on RF access links, can be reformulated as the following equivalent form
\begin{subequations}\label{eq:optSDPrankone}
	\begin{align}
		(\textbf{OPT-3})\!:\quad&\underset{\{ \mathbf{W}_{k,c,t} \in\mathbb{C}^{M\times M}\}}
		{\text {minimize}}~ \sum\limits_{k=1}^K \sum\limits_{c=1}^C \omega \Tr (\mathbf{W}_{k,c,t})
		\label{eq:ObjFunBF1}\\
		\text {s.t.} \qquad \qquad
		& \hspace{-2.85em} \frac{\Tr (\mathbf{W}_{k, c, t}{\bf H}_{k,i,t})}{\delta_c} 
		\geq 
		\sigma_{RF}^2 + \sum\limits_{l\neq c}^{C} \Tr( {\bf W}_{k,l,t} {\bf H}_{k,i,t} ), \notag \\
		& \hspace{8.85em} \ \forall i \in  \mathcal{G}_{k,c}^t, \forall c, \forall k, \label{eq:SDPQosConstr}\\
		& {\bf W}_{k,c,t}\succeq 0,~\forall c, \forall k,\label{eq:SDPConstr}\\
		& \rank({\bf W}_{k,c,t}) = 1, ~\forall c, \forall k,\label{eq:RankOne}
	\end{align}
\end{subequations}
where ${\bf H}_{k,i,t} = {\bf h}_{k,i,t}{\bf h}_{k,i,t}^H$. 
Note that $\Tr (\mathbf{X}_{k,c,t}) = \|\mathbf{w}_{k,c,t}\|_2^2$ if and only if $ {\bf W}_{k,c,t}\succeq 0$ and $\rank({\bf W}_{k,c}) = 1$ \cite{Mehanna2013}.
 
By dropping the non-convex rank-one constraints \eqref{eq:RankOne}, the problem (\textbf{OPT-3}) can be relaxed to the following convex semi-definite programming (SDP) problem \cite{Karipidis2008}
\begin{subequations}\label{eq:optSDP}
	\begin{align}
		(\textbf{OPT-4})\!:\quad&\underset{\{ \mathbf{W}_{k,c,t} \in\mathbb{C}^{M\times M}\}}
		{\text {minimize}}~ \sum\limits_{k=1}^K \sum\limits_{c=1}^C \omega  \Tr (\mathbf{X}_{k,c,t})
		\label{eq:ObjFunBF}\\
		\text {s.t.} \qquad \qquad
		&  \eqref{eq:SDPQosConstr}-\eqref{eq:SDPConstr}.
	\end{align}
\end{subequations}

As a standard SDP problem, (\textbf{OPT-4}) can be efficiently solved in polynomial time using interior-point methods \cite{Boyd2004}.
However, due to the rank relaxation, the solution of (\textbf{OPT-4}), denoted by $\{\mathbf{W}_{k,c,t}^*\}$, may include matrices that are not rank-one.
If the derived solution contains only rank-one matrices, the set of optimal beamforming vectors $\mathcal{W}^*_t=\{\mathbf{w}_{k,c,t}^*|\forall k, \forall c\}$ for problem (\textbf{OPT-1}) can be obtained by performing eigenvalue decomposition on $\{\mathbf{W}_{k,c,t}^*\}$.
Otherwise, we generate a set of candidate beamforming matrices based on $\{\mathbf{W}_{k,c,t}^*\}$ using the Gaussian randomization technique and then select the one that minimizes the power cost while ensuring feasibility \cite{Mehanna2013}.

\begin{algorithm}[!htb]
	\DontPrintSemicolon
	\caption{PPO-Based Algorithm}\label{alg:PropAlg}
	Initialize $\theta, \phi$ randomly\\
	\For{ ${i}_{\text{outer}}=1,2,\dots, Iter_{\max}$}{
		\For{ $t=0,2,\dots,T-1$}{
			Run policy $\pi_{\theta_{{i}_{\text{outer}}}}$ to generate action ${\bf a}_t$ \;
			Calculate $\{\mathcal{T}^*_t, \mathcal{E}^*_t\}$ by solving problem (\textbf{OPT-3}) \;
			Calculate $\mathcal{W}^*_t$ by solving problem (\textbf{OPT-4}) \;
			Store transistion sample $ ( {\bf s}_t, {\bf a}_t, r_t = -q^*_t, {\bf s}_{t+1} ) $   
		}
		Compute advantage estimation $\hat{A}(s_n,a_n)$ by \eqref{eq:gae} \;
		Compute target value $\hat{V}(s_n)$ by \eqref{eq:cal_tar} \;
		\For{ ${i}_{\text{inner}}=1,\dots, Iter_{\rm MB} $}{
			Randomly sample a mini-batch of size \( N_B \) from collected transitions \;
			Update  parameters \( \theta \) and \( \phi \) according to \eqref{eq:update_theta} and \eqref{eq:update_phi} \;
		}
	}
\end{algorithm}

The pseudocode for our proposed algorithm is provided in Algorithm \ref{alg:PropAlg}.
The overall complexity of Algorithm \ref{alg:PropAlg} is impacted by the complexity of solving the optimization problems (\textbf{OPT-3}) and (\textbf{OPT-4}) as well as the parameter update steps for \( \theta \) and \( \phi \).
The worst-case computational complexity for solving a convex problem $N_{{\rm v}}$ variables and $N_{{\rm c}}$ constraints using interior point methods is given by $\mathcal{O}\bigl(Iter_{{\rm IPM}} \max(N_{{\rm v}}^3,  N_{{\rm v}}^2 N_{{\rm c}} )\bigr)$, where $Iter_{{\rm IPM}}$ denotes the maximum number of iterations required by the interior point method \cite{Boyd2004}.
Therefore, the computational complexity of solving problems (\textbf{OPT-2}) and (\textbf{OPT-4}) can be expressed as
$\mathcal{O}\bigl( Iter_{{\rm IPM}} \max\bigl( (CKL +LC + 2L)^3, (CKL +LC + 2L)^2(2CK + CKL +2L+D+K) \bigr) \bigr)$ 
and 
$\mathcal{O}\bigl(Iter_{{\rm IPM}} \max\bigl(K^3C^3M^6,  K^2C^2M^4(KC + U) \bigr)\bigr)$.
Meanwhile, according to~\cite{goodfellow2016deep}, the time complexity of updating the parameterized 
neural networks $\pi_{\theta}$ and $V_{\phi}$ can be expressed as $\mathcal{O}\bigl(\sum\limits_{i=1}^{N_L^{\theta}-1} n^\theta_i n^\theta_{i+1}\bigr)$ and  $\mathcal{O}\bigl(\sum\limits_{i=1}^{N_L^{\phi}-1} n^\phi_i n^\phi_{i+1}\bigr)$, respectively. 
Here, $n^\theta_i$ and $n^\phi_i$ represent the number of neurons in the $i$-th layer of 
$\theta$ and $\phi$.
$N_L^{\theta}$ refers to the total number of layers in the network associated with $\theta$, and $N_L^{\phi}$ refers to the number of layers in the network associated with $\phi$.
Specifically, the number of neurons in the first layer of both $\pi_{\theta}$ and $V_{\phi}$ corresponds to the state dimension $\dim({\bf s}_t)=2KC$, while the number of neurons in the last layer of  $V_{\phi}$ is $1$,
and in $\pi_{\theta}$, it equals the action dimension $\dim({\bf a}_t)=KC$.
Given that the outer loop runs for $Iter_{\max}$ iterations and the inner loop (for parameter updates) runs for  $Iter_{{\rm MB}}$ iterations,
the overall complexity for the entire algorithm is  
$\mathcal{O}
\biggl(
Iter_{\max}
\Bigl( 
	T \bigl( Iter_{{\rm IPM}} \max\bigl( (CKL +LC + 2L)^3, (CKL +LC + 2L)^2(2CK + CKL +2L+D+K) 
			+ Iter_{{\rm IPM}} \max\bigl(K^3C^3M^6,  K^2C^2M^4(KC + U) \bigr)
	  \bigr)
	+
	Iter_{{\rm MB}} 
	\bigl( 
	\sum\limits_{i=1}^{N_L^{\theta}-1} n^\theta_i n^\theta_{i+1} + 
	\sum\limits_{i=1}^{N_L^{\phi}-1} n^\phi_i n^\phi_{i+1}
	\bigr) 
\Bigr)
\biggr)$.

\section{Numerical Results}\label{sec:sim_res}
In this section, we evaluate the performance of our proposed algorithm through comprehensive simulations.
We consider a network with $D=2$ DCs, $K=7$ HAPs, and $M=6$ RF antennas at each HAP. 
t is assumed that each HAP provides coverage over a circular area with a radius of 15 {\rm km}, and the horizontal distances between the centers of any two adjacent HAPs range from 40 {\rm km} to 60 {\rm km}.
The parameters chosen in this section are based on existing literature such as \cite{Liu2024a,Farid2007, Bashir2023, Barrios2012, Breslau1999}.
The channel coefficient between the antenna of the HAP and the user is modeled as the product of free-space path loss and Rician small-scale fading with a factor of $\kappa_{{\rm RF}} = 5$ \cite{Liu2024a}.
The wavelength of the FSO signal is $1550~{\rm nm}$, the responsivity of the detector is $\varrho = 0.6$, the standard deviation of the angular pointing error is $\sigma_0 = 20~{\rm mrad}$, the angular beamwidth is $\vartheta = 40~{\rm mrad}$, and the aperture radius is $\chi_{\rm d} = 0.4~{\rm m}$ \cite{Farid2007, Bashir2023}. 
The parameters of the exponentiated Weibull distribution accounting for the atmospheric turbulence loss of FSO links are $\varphi=3.21$, $\varsigma=1.25$, and $\varepsilon = 0.94$ \cite{Barrios2012}.  
Without loss of generality, we set $\mu_c^{\rm acc} = \mu^{\rm acc}$ and $\mu_c^{\rm cac} = \mu^{\rm cac}$, $\forall c$.
Unless specified otherwise, we consider $U=105$ terrestrial users, and each user submits a file request independently to a database of $C=30$ contents, $\omega = 1$, $B_{\rm FSO} =10~{\rm GHz}$, $B_{\rm RF} =10~{\rm MHz}$, $\mu^{\rm cac} = 10~{\rm Mbps}$, $\mu^{\rm acc} = 4~{\rm Mbps}$, and the popularity of contents for users covered by different HAPs in each time slot follows independent Zipf distributions with different skewness parameters in the range from $0.5$ to $4$ \cite{Breslau1999}.
All the hidden layers in DNNs use the $\FuncSty{Tanh}$ activation and the hyper-parameters utilized in the DRL model are detailed in Table~\ref{Tab:hp_drl}. 

\begin{small}
	\begin{table}[!htb]
		\caption{Hyper-Parameters of DRL}
		\centering
		\begin{tabular}{c|c}
			\hline\hline
			\textbf{Parameter} &  \textbf{Value}\\
			\hline
			Number of hidden layers (all neural networks) & 2 \\
			\hline
			Number of neurons in hidden layers & [256, 128]\\
			\hline
			Mini-batch size & 32 \\
			\hline
			Discount factor & 0.99\\
			\hline
			Actor Module learning rate & 3e-4\\
			\hline
			Critic Module learning rate & 3e-4\\
			\hline
			Soft update learning rate & 5e-3\\
			\hline
			SGD optimizer & $\mathrm{Adam}$~\cite{kingma2014adam}\\
			\hline \hline
		\end{tabular}
		\label{Tab:hp_drl}
	\end{table}
\end{small}

\subsection{Effect of Weighting Parameter}

\begin{figure}[!htb] \centering \centering
	\begin{minipage}{1\linewidth}
		\centering
		\subfloat[]{\label{fig:per_cov}\includegraphics[width=0.8\linewidth]{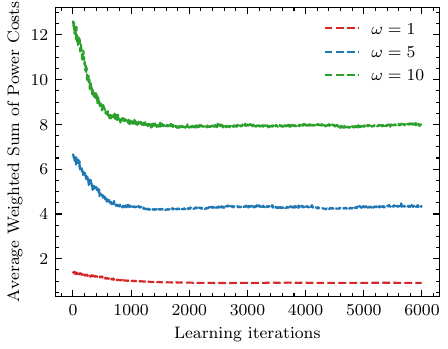}}
	\end{minipage}%
	\hspace{1mm}
	%\hfill
	\begin{minipage}{1\linewidth}
		\centering
		\subfloat[]{\label{fig:pwrHAP}\includegraphics[width=0.8\linewidth]{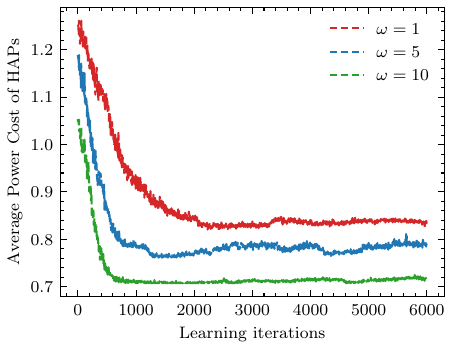}}
	\end{minipage}
	\hspace{1mm}
	%\hfill
	\begin{minipage}{1\linewidth}
		\centering
		\subfloat[]{\label{fig:pwrDC}\includegraphics[width=0.8\linewidth]{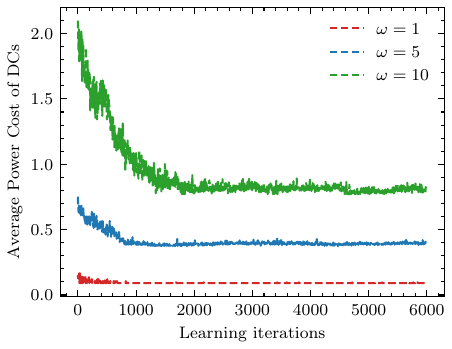}}
	\end{minipage}\par\medskip
	\caption{Performance of Algorithm \ref{alg:PropAlg} with respect to the number of learning iterations: (a) Average weighted sum of power cost, (b) Average power consumption of HAPs, (c) Average power consumption of DCs.}
	\label{fig:combined}
\end{figure}

The effectiveness of the proposed DRL and convex optimization approach is demonstrated by tracking the evolution of power consumption across learning iterations for different values of the weighting parameter $\omega$.
\figref{fig:per_cov} presents the weighted sum of the power costs, which serves as the objective function in our optimization. 
It is evident that the weighted sum of the power costs decreases as the learning progresses.
Moreover, assigning a higher weight to the HAPs' power consumption leads to a higher weighted sum of the power costs.
As shown in \figref{fig:pwrHAP}, the average power consumption of HAPs steadily decreases as $\omega$ increases.
Conversely, the power consumption of DCs in \figref{fig:pwrDC} follows an inverse pattern, and the variation between different $\omega$ values is less pronounced compared to that of HAP power consumption.
These results confirm that the proposed optimization framework effectively adapts to different weight settings, yielding efficient power management while balancing the contributions of DCs and HAPs.
To verify the effectiveness of our proposed method, we next compare it with the following baseline schemes. 
\begin{itemize}
	\item Baseline 1: This approach differs from our proposed method in that the backhaul traffic is transmitted in a unicast fashion, and the caching decision is made by a DRL agent. 
	\item Baseline 2: Each HAP caches the most popular content until its storage is full, based on perfect knowledge of forthcoming content access requests from its covered users, while resource allocation on the backhaul and access links follows the same strategy as in our proposed method.
	\item Baseline 3: Each HAP randomly caches content with equal probabilities, regardless of popularity distribution, and resource allocation on the backhaul and access links is identical to that in our proposed method.
	\item Baseline 4: In this scheme, HAPs have no caching capability.
	Consequently, all content requests must be fulfilled by retrieving data directly from the data centers via the backhaul network. The resource allocation strategy for both the backhaul and access links remains consistent with the approach used in our proposed method.
\end{itemize} 

\subsection{Effect of Number of Contents} % 

\begin{figure}[!htbp]
	\center
	\includegraphics[width=0.8\columnwidth]{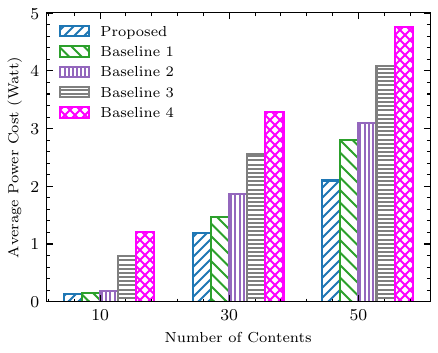}
	\caption{Power cost versus number of contents.}
	\label{fig:barNumConts}
\end{figure}

\figref{fig:barNumConts} illustrates the power cost under different number of contents. 
The proposed method consistently achieves the lowest average power consumption across all content quantities, outperforming all considered baselines. 
Baseline 1, which utilizes unicast transmission and DRL-based caching decisions, exhibits higher power consumption that increases with the number of contents.  
While Baseline 2 employs network coding–based multicast transmission in the backhaul links and caches the most popular content based on perfect knowledge of future requests, it still lags behind both the proposed method and Baseline 1 in power consumption.
This demonstrates that a DRL-based proactive caching strategy is indeed crucial for improving energy efficiency.
Baseline 3 employs random caching and incurs significantly higher power costs compared to our proposed method as well as Baselines 1 and 2.
Baseline 4, where the HAPs are unable to cache content, leads to the highest power consumption across all content quantities.
These findings underscore the effectiveness of our proposed method in minimizing power cost, particularly as the number of contents increases.

\subsection{Effect of Number of Users} % 10; 15; 20

\begin{figure}[!htbp]
	\center
	\includegraphics[width=0.8\columnwidth]{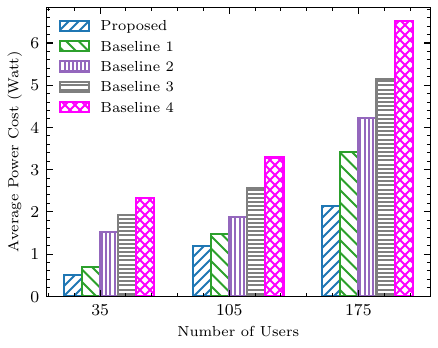}
	\caption{Power cost versus number of users.}
	\label{fig:barNumUrs}
\end{figure}

\figref{fig:barNumUrs} shows the changes of power cost with respect to the number of users. 
The average power cost increases as the number of users increases, since increasing the number of users subsequently increases the traffic load over the network. 
Compared to the baselines, our proposed algorithm achieves at least a $19.4\%$ improvement in average power cost across all considered scenarios.
Specifically, when $U=175$, it is shown in \figref{fig:barNumUrs} that our proposed algorithm outperforms other considered baselines by at least $37.7\%$ in terms of average power cost.
These results highlight the superior efficiency of the proposed method and demonstrate that joint optimization in both caching strategy and resource allocation is crucial for handling the incremental service demand resulting from increasing population density.

\subsection{Effect of Data Rate Requirements}\label{sc:EDRR}

\begin{figure}[!htb] \centering \centering
	\begin{minipage}{0.8\linewidth}
		\centering
		\subfloat[]{\label{fig:barCacRateReq}\includegraphics[width=\linewidth]{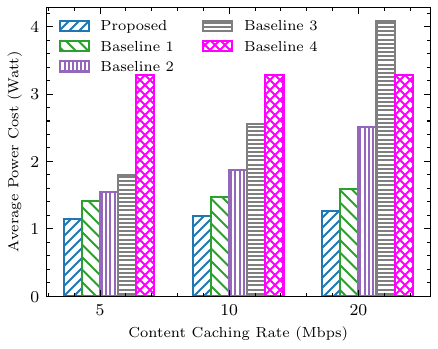}}
	\end{minipage}%
	\hspace{1mm}
	\begin{minipage}{0.8\linewidth}
		\centering
		\subfloat[]{\label{fig:barAccRateReq}\includegraphics[width=1\linewidth]{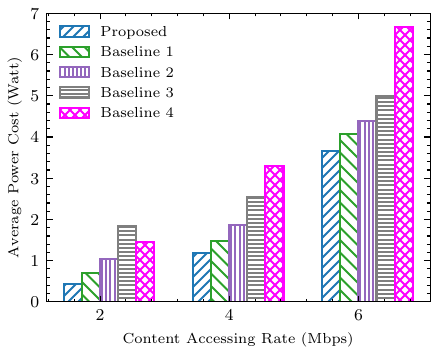}}
	\end{minipage}%
	\hspace{1mm}
	\caption{Effect of the data Rate Requirements: (a) content caching, (b) content accessing.}
	\label{fig:barRateReq}
\end{figure}

\figref{fig:barCacRateReq} and \figref{fig:barAccRateReq} demonstrate the impact of content caching and content accessing data rate requirements on average power cost, i.e., $\mu^{\rm cac}$ and $\mu^{\rm acc}$, respectively. 
Except for Baseline 4, it is evident that as the rate requirement for either content caching or content access increases, the average power cost rises across all considered schemes.
This outcome is expected as a higher data rate requirement within the same transmission period necessitates higher transmitting power. 
Since Baseline 4 lacks caching capabilities, increasing the data rate requirement for content caching has no effect on its performance.
In addition, the impact of the content accessing data rate requirement is more pronounced than that of the content caching data rate requirement, as the content caching data rate only affects backhaul transmission, whereas the content accessing data rate influences both the access links and the backhaul link when the HAP cannot directly fulfill the content access requests from its covered users.
Our proposed method consistently exhibits the lowest average power cost in all considered cases, with up to $69.1\%$ and $76.8\%$ performance improvements under the specified content caching and content accessing data rate requirements, respectively.  

\subsection{Effect of Local Cache Capacity on the HAP} % 

\begin{figure}[!htbp]
	\center
	\includegraphics[width=0.8\columnwidth]{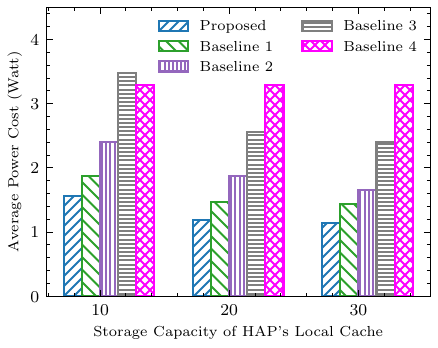}
	\caption{Average power versus local cache storage capacity on the HAP.}
	\label{fig:barStoCap}
\end{figure}
In \figref{fig:barStoCap}, we evaluate the performance of power cost under varying storage capacities of local cache at the HAP.
For the same reason discussed in Section \ref{sc:EDRR}, the average power cost for Baseline 4 remains unchanged in \figref{fig:barStoCap}.
It is worth noting that the performance improvement for both our proposed method and most baselines is less significant when increasing from $N^{{\rm sto}}=10$ to $N^{{\rm sto}}=20$ compared to the increase from $N^{{\rm sto}}=20$ to $N^{{\rm sto}}=30$,  which suggests that simply expanding the local cache storage capacity is not an efficient way to reduce power cost. 
When $N^{\rm sto}=10$, Baseline 3 performs worse than Baseline 4, indicating that a random caching strategy may be less effective than not caching at all.
This emphasizes the necessity of proper caching design, as demonstrated by our proposed method, which achieves an average $34.8\%$ improvement in power cost performance across all considered local cache capacity conditions.

\subsection{Effect of Channel Bandwidth}

\begin{figure}[!htb] \centering \centering
	\begin{minipage}{0.8\linewidth}
		\centering
		\subfloat[]{\label{fig:barFSOChnBanw}\includegraphics[width=\linewidth]{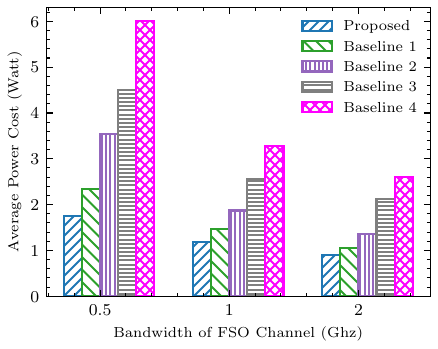}}
	\end{minipage}%
	\hspace{1mm}
	\begin{minipage}{0.8\linewidth}
		\centering
		\subfloat[]{\label{fig:barRFChnBanw}\includegraphics[width=1\linewidth]{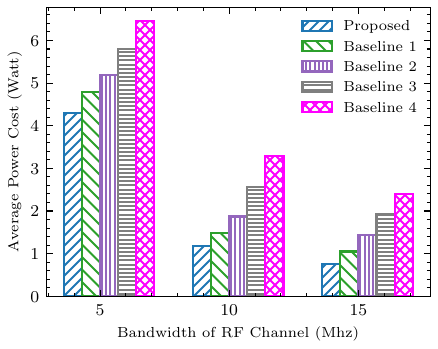}}
	\end{minipage}%
	\hspace{1mm}
	\caption{Effect of the channel bandwidth: (a) FSO, (b) RF.}
	\label{fig:ChnBanw}
\end{figure}

\figref{fig:ChnBanw} plots the power cost under different FSO backhaul and RF access channel bandwidths.
We can observe from \figref{fig:ChnBanw} that the average power cost of all schemes decreases as the FSO channel or RF channel bandwidth increases, and our proposed method achieves the lowest average power post in all scenarios.
In particular, our proposed algorithm yield average $45.8\%$ and $38.6\%$ power cost performance improvements for considred setting of FSO channel bandwidth and RF channel bandwidth, respectively.  
\figref{fig:barFSOChnBanw} and \figref{fig:barRFChnBanw} both show that the marginal benefits of continuously increasing channel bandwidth become less significant. 
However, the impact of RF channel bandwidth is slightly more pronounced at the largest bandwidth value compared to the mid-range.
Moreover, as illustrated in \figref{fig:barFSOChnBanw}, the network coding-based routing solution is more effective when the backhaul channel bandwidth is limited.

\subsection{Effect of Visibility Condition}\label{sec:new_resu} % 
\begin{figure}[!htbp]
\center
\includegraphics[width=0.8\columnwidth]{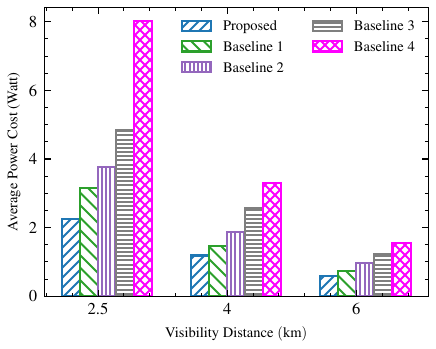}
\caption{Average power versus visibility condition.}
\label{fig:barVisDist}
\end{figure}
In \figref{fig:barVisDist}, we plot the performance of average power cost under different visibility parameters $\nu$ corresponding to various weather conditions. 
In particular, we examined three atmospheric scenarios: thick haze with visibility of 2.5 {\rm km}, light haze with visibility of 4 {\rm km}, and near-clear conditions with visibility of 6 {\rm km}, to evaluate the impact of varying weather conditions and their associated atmospheric attenuation on signal propagation and power requirements.
It is evident that adverse weather conditions  significantly increase power demands due to reduced visibility and higher attenuation. Conversely, as visibility distance increases under improved weather conditions, the average power cost decreases.
As shown in \figref{fig:barVisDist}, even in the worst visibility condition, our proposed mechanism can achieve at least  $28.9\%$  performance improvement over baseline methods, demonstrating robustness against weather-driven atmospheric degradation.

\section{Conclusion}
In this work, we tackled the challenge of energy-efficient content delivery in rural areas by utilizing multi-HAP networks equipped with FSO backhaul and RF access links.
Our proposed hierarchical framework, which integrates deep reinforcement learning and convex optimization, effectively optimizes the dynamic caching strategy and resource allocation to minimize long-term power cost. 
By incorporating network coding-based multicasting, our approach further enhances network efficiency by treating different content types as distinct multicast sessions. 
Through extensive simulations, we demonstrated that the proposed method significantly reduces power consumption compared to existing baseline approaches. These results indicate that our framework offers a practical solution for bridging the digital divide and improving connectivity in underserved rural regions. 

\bibliographystyle{IEEEtran}
\bibliography{IEEEabrv,hokie-HD}
\end{document}